\documentclass{iopart}  
\usepackage{epsfig}

\newcommand{\be}[1]{\begin{equation}\label{#1}}
\newcommand{\ee}{\end{equation}}   
\newcommand{\bea}{\begin{eqnarray}}
\newcommand{\eea}{\end{eqnarray}} 
\newcommand{\barr}{\begin{array}}
\newcommand{\earr}{\end{array}}
\newcommand{\eq}[1]{(\ref{#1})}

\newcommand{\PRA}{{\it Phys.~Rev.~A} }

\newcommand{\showlabel}[1]{}

\begin{document}

\title[Classical electron impact double ionization of helium]
{Electron impact double ionization of helium from classical trajectory
calculations}

\author{Tiham\'{e}r Geyer%
	\footnote[1]{Present address: Zentrum f\"ur Bioinformatik, 
		Universit\"at des Saarlandes, D--66041 Saarbr\"ucken, Germany}
}

\address{Department of Chemical Physics, Weizmann Institute of Science, 
		Rehovot 76100, Israel}

\begin{abstract}
	With a recently proposed quasiclassical ansatz [Geyer and Rost,
	\emph{J. Phys.  B} \textbf{35} (2002) 1479] it is possible to
	perform classical trajectory ionization calculations on many
	electron targets.  The autoionization of the target is prevented by
	a M\o{}ller type backward--forward propagation scheme and allows to
	consider all interactions between all particles without additional
	stabilization.  The application of the quasiclassical ansatz for
	helium targets is explained and total and partially differential
	cross sections for electron impact double ionization are
	calculated.  In the high energy regime the classical description
	fails to describe the dominant TS1 process, which leads to big
	deviations, whereas for low energies the total cross section is
	reproduced well.  Differential cross sections calculated at 250 eV
	await their experimental confirmation.
\end{abstract}


\submitto{\jpb}


\section{Introduction}

Classical trajectory calculations of scattering processes have many
advantages: they are easily implemented and robust and they provide an
interpretation in terms of moving point particles, which is very
familiar to our daily life's experience \cite{OLS03}.  But despite all
their ``beauty'' and simplicity they have two serious drawbacks: they
often are far from accurate and, what is even more limiting, they are
conceptually restricted to one electron atoms --- many electron atoms
``autoionize''.

In a classical method, which starts from Bohr's action quantization
\cite{ABR66}, only the hydrogen atom is stable.  Actually this is the
correct behavior: classical mechanics is the limit of very high quantum
numbers.  Consequently a classical hydrogen atom corresponds to a
highly excited Rydberg state, which can only decay radiatively.  It is
therefore stable in the classical treatment.  But a classical two
electron atom resembles a doubly excited helium atom, which is only
metastable --- the ``annoying'' autoionization of a classical many
electron atom is the correct physical behavior.  To use such an
unstable atom as a target in a classical trajectory calculation despite
this difficulty two modifications of the Hamiltonian have been
proposed: either the electron--electron interaction is neglected in the
independent electron model (IEM) or, as in the ``nCTMC'' ansatz,
additional stabilization potentials are included, which model the
uncertainty relation \cite{KIR80} or place a bound on the accessible
binding energies \cite{COH96}.  But these amendments are quantal
additions onto the otherwise purely classical ansatz.

The stability problem might still be called a technical problem, but
the fundamental difficulty remains: Bohr's action quantization is a one
dimensional rule only.  It works for hydrogen, as this problem
factorizes in spherical coordinates; each degree of freedom is
quantized on its own.  For the helium atom there exists no coordinate
system in which all degrees of freedom separate.  This is connected to
the fact that some quantum numbers have to be replaced by ``propensity
rules'' in a quantal description of the helium atom \cite{TAN00}.  In
the context of semiclassical calculations some multi dimensional
quantization rules have been proposed, but they are limited to special
geometries \cite{ZAK88,BOG92}.  Consequently there exists no
prescription that would tell how to extend Bohr's hydrogen model into a
classical helium atom.

For the hydrogen target, where these two problems do not exist, the
cross sections from a classical trajectory calculation are often not
good --- and therefore never were published --- especially when angle
differential cross sections are considered. The obvious way to try to
improve the results is to modify the phase space description of the
hydrogen target by including quantum mechanical information. This
works to some extent, but leads either to unstable targets like the
Wigner distribution \cite{EIC81,COH85} or to the need for fit
parameters \cite{HAR83}. Also with these alternate descriptions the
initial values are not confined to the energy shell any more, an
inconsistency that was neglected when extracting cross sections.

In a recent work we showed on electron impact ionization of atomic
hydrogen how these fundamental problems can be overcome and the cross
sections improved by a rederivation of the classical approximation of
the time dependent quantum mechanical treatment \cite{GEY02}: the
scattering process is first formulated quantum mechanically in the
M\o{}ller picture and then all parts --- the initial state, the
scattering operator in its time dependent form and the final state ---
are simultaneously translated into Wigner's formulation of quantum
mechanics and only then approximated classically by setting $\hbar =
0$. The backward--forward structure of the resulting propagation
scheme, a consequence of the M\o{}ller form of the scattering
operator, refocuses the decaying classically approximated target onto
the collision event. The energy spread of the initial state is dealt
with by switching to cross sections differential in the energy
transfer instead of using the absolute final energies of each
trajectory. With this approach the cross sections on hydrogen could be
improved greatly over a wide range of energies and geometries compared
to the standard CTMC procedure. Even fully differential cross sections
could be reproduced.

This quasiclassical approach now allows for arbitrary target
descriptions, which do not need to be stable and need not be confined
to the energy shell.  There is also no fundamental limit on the
dimensionality of the initial state, i.e., on the number of target
electrons: the initial distribution to be used in the classical
calculation is derived from the wave function and not from torus
quantization according to quantum numbers, so the way to use multi
electron targets in a CTMC like method is open.  In a recent letter
\cite{GEY03} we demonstrated that this dynamical stabilization through
the propagation scheme in fact makes it possible to perform ionization
calculations on classical helium targets.  We calculated total single
\emph{and} double ionization cross sections on helium simultaneously
without any modification of the interactions between all particles
involved.

In this paper we will present further details of the calculation and
show and interpret differential cross sections for electron impact
double ionization of atomic helium.

\section{The quasiclassical procedure}

As mentioned above the quasiclassical approach is a classical 
approximation to quantum mechanical scattering, in which the quantum 
formulation is translated into phase space through the use of a
correspondence rule \cite{MOY49,COH66} and then approximated 
with $\hbar=0$. This procedure was presented in detail in 
\cite{GEY02}, so we just give a short summary and put the emphasis on 
these parts, which have to be adapted for multi electron targets.

\subsection{Scattering operator and M\o{}ller propagation scheme}

The backbone of the description is the scattering
operator $\hat{S}$ in the M\o{}ller formulation (see, e.g.
\cite{TAY72}):
\be{}
	\hat{S} = \Omega^\dag_- \Omega_+ 
	\quad \mbox{with} \quad
	\Omega_\pm = \lim_{t \to \mp \infty} U^\dag(t) U_0(t),
\ee
the M\o{}ller operators. The propagators $U(t) = \exp[-iHt]$ and
$U_0(t) = \exp[-iH_0t]$ are defined in the usual way as the formal
solution of the time dependent Schr\"o{}dinger equation $\dot{\psi} =
-\rmi H \psi$ and describe the time evolution of $\psi$ under the full
and the asymptotic Hamiltonian, respectively.

This ansatz partitions the scattering process into three distinct 
stages, which are characterized by the considered interactions between 
the particles. The first and the last asymptotic propagation stages 
serve to construct the scattering states, which are solutions of the 
full Hamiltonian $H$, from the asymptotic initial and final states, 
which in turn are eigenfunctions of $H_0^i$ and $H_0^f$, respectively. 
For electron impact ionization of atomic helium the full Hamiltonian 
has the following form:
\be{eq:Hamiltonian}
	H = \frac{p_p^2}{2} - \frac{Z}{r_p} + \frac{p_1^2}{2} - 
		\frac{Z}{r_1} + \frac{p_2^2}{2} - \frac{Z}{r_2} + 
		\frac{1}{|r_p-r_1|} + \frac{1}{|r_p-r_2|} + 
		\frac{1}{|r_1-r_2|}\quad
\ee
$H$ is
given in atomic units, which we will use throughout the paper, if not
otherwise denoted. The subscript $p$ denotes the projectile, while the
target electrons are labelled with 1 and 2. $Z=2$ is the charge of the
target nucleus, which we assume to have an infinite mass. In the
asymptotic initial channel, described by
\be{eq:HInitial}
	H_0^i = \frac{p_p^2}{2} + 
		\frac{p_1^2}{2} - \frac{Z}{r_1} + 
		\frac{p_2^2}{2} - \frac{Z}{r_2} + 
		\frac{1}{|r_1-r_2|}\, ,
\ee
the projectile electron and the neutral helium target are independent, 
whereas in the asymptotic final state the ejected electrons do not 
interact with each other and the residual ion. We then have to 
distinguish between single and double ionization and we consequently 
have to consider two different asymptotic final Hamiltonians 
$H_0^{f(1)}$ and $H_0^{f(2)}$:
\bea
	H_0^{f(1)} & = & \frac{p_a^2}{2} + \frac{p_b^2}{2} +
		\frac{p_c^2}{2} - \frac{Z}{r_c} 
		\label{eq:HFinal1}\\
	H_0^{f(2)} & = & \frac{p_a^2}{2} + \frac{p_b^2}{2} +
		\frac{p_c^2}{2}
		\label{eq:HFinal2}
\eea
\showlabel{eq:HFinal1}\showlabel{eq:HFinal2}
The electrons are indistinguishable after the scattering event, we
therefore label them with the subscripts $a$, $b$ and $c$ in the final
state, where $a$ has the highest and $c$ the lowest final energy.

In the classical limit the M\o{}ller procedure translates into a
propagation scheme, where each trajectory is first propagated
backwards in time with $H_0^i$ and then forward under $H$, i.e., with
all interactions on. At first this may seem like superfluous effort,
but this is the crucial difference to standard CTMC. It allows to use
initial state distributions, which are not stationary under the
approximated classical propagation: in the case of two electrons and a
nucleus --- the helium atom --- one of the electrons will be ejected
already in the initial backward propagation. When the propagation is
reversed this ejected target electron is ``brought back'' into the
atom. So when the projectile encounters the target both electrons have
essentially returned their initial position. Because of this
refocussing by the M\o{}ller propagation scheme there is no more need
for any additional stabilization potentials, although all interactions
are included correctly.

In a sense time has been reduced to a mere integration variable, as in
the M\o{}ller formulation all the important events --- the preparation
of the initial state, the actual scattering process and the extraction
of the cross sections --- happen at the same ``position'' in time
$t=0$.

A different way to view the effect of the backward--forward 
propagation is the following: the classical approximation to the 
correct Wigner propagation neglects all terms containing $\hbar$, so 
errors are introduced. But, neglecting the approaching projectile for 
a moment, after going back and forth in time for the same interval 
the system returns to the same state again: the combined propagation 
in both directions gives the same final result as the correct Wigner 
propagation, only the intermediate steps, for which $t<0$, are 
different. With the projectile added this is not completely true any 
more, but most of the error of the classical approximation is 
cancelled, when it comes to the target electrons.

\subsection{The initial state distribution}
\label{sec:InitialState} \showlabel{sec:InitialState}
 
In our quasiclassical approximation the initial distribution is
constructed from the quantum mechanical wave function, which describes
the initial state according to the asymptotic initial $H_0^i$.  With
our choice \eq{eq:HInitial} this is a product wave of the free
projectile electron and the helium target in its ground state.  These
two independent wave functions are translated into the Wigner picture
\cite{WIG32} and then the product of their classical $\hbar=0$
approximations is discretized, each discretization point serving as
initial conditions for a trajectory in the above explained
backward--forward propagation scheme. 

Due to the $\hbar=0$ approximation each of these trajectories evolves
independently, but that does not imply that they are to be interpreted
independently of each other.  As the initial state is modelled by all
initial values collectively, also their entirety only describes the scattering
process.  Remeber that we still deal with a description in wave
functions, though obscured by the classical approximation and the
discretization.

For helium and more complex atoms there arises an additional
difficulty compared to hydrogen: there the exact analytic wave
function is known, whereas for helium only approximate wave functions
of varying complexity have been formulated
\cite{SLA27,HYL28,HYL29,PEK58,KIN59}.  We therefore have to choose not only
a prescription for how to translate the quantum wave function into a
classical distribution \cite{GEY02,MOY49,COH66,WIG32} but also an
appropriate wave function to start from.

For this first analysis we started from the most simple quantum
mechanical ansatz for the helium wave function. We use the ``$1s^2$''
product state of two identical hydrogenic ground state wave functions
with an effective nuclear charge $C$. This product wave
$\psi(r_1,r_2)=$ H1s$_C(r_1)$ H1s$_C(r_2)$ is then translated into a
product of two identical classical phase space distributions.
\be{eq:HeWF}
	\psi(r_1, r_2) = \frac{C^3}{\pi} \exp(-Cr_1) \exp(-Cr_2)
\ee
Independent of the nuclear charge this product wave function yields
the minimum energy at $C=Z-5/16$ \cite{HYL28}, i.e., $C=27/16$ for
helium. This results in an energy of $E_t=-2.85$ a.u. ($-$77.5 eV),
slightly less than the experimental value of $-2.904$ a.u. ($-$79 eV).

The angular electron--electron correlation is neglected in the initial
wave function \eq{eq:HeWF}. Nevertheless we include the mutual
repulsion of the target electrons in $H_0^i$, so their motion is
coupled from the very moment the propagation is started. Of course,
exchange correlation effects are beyond the reach of a classical
approximation.

As shown by Cohen \cite{COH66} a generalization of Weyl's
correspondence rule allows to derive nearly arbitrary recipes to set up
the initial distribution within this quasiclassical ansatz.  Each of
them has different properties, which are important in different energy
regimes, as was shown in \cite{GEY02}.  We use three of the
prescriptions described there, to convert the one electron wave
functions H1s$_C$ of equation \eq{eq:HeWF} into phase space
distributions: (i) the ``product'' distribution \cite{COH66}, where the
single electron wave function is translated into the product of the
coordinate and momentum space densities, (ii) the ``microcanonical''
distribution \cite{ABR66}, where each of the electrons is sampled
according to a microcanonical distribution at an adapted one electron
energy of $-1.94$ a.u., so that the (averaged) total energy, including
the electron's interaction, resembles that of the helium ground state
and (iii) ``Cohen's energy distribution'' \cite{COH96}, which is a
superposition of microcanonical distributions with an energy dependent
weight function, rescaled for the effective $C$.

To avoid clumsy wording we will label these distributions in the 
following as ``product distribution'', ``microcanonical distribution'' 
and ``Cohen's (energy) distribution'', respectively, though these 
names are originally associated with the underlying one electron 
distributions.

In the calculations on hydrogen \cite{GEY02} the product distribution
performed best for high energies, whereas Cohen's energy distribution
gave the best results at very low total energies. These two reproduce
the density of the hydrogen ground state both in coordinate and
momentum space. The microcanonical distribution, which is the only
purely classically derivable distribution, has, due to its fixed
binding energy, the wrong spatial density and gave worse results for
all energies considered than any of the other distributions.

\epsfxsize=8cm
\epsfysize=5cm
\begin{figure}[t]
	\centerline{\epsfbox{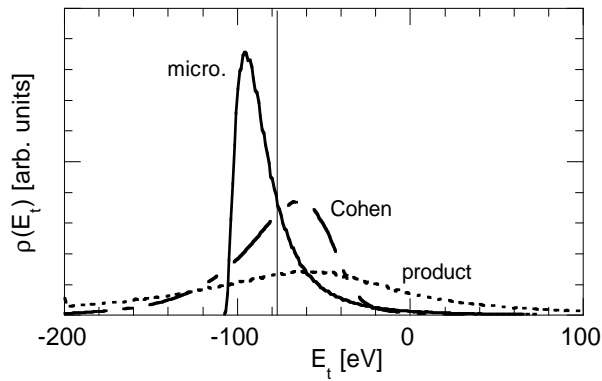}}
	\caption{Distribution of the target energy when the initial
	conditions are chosen according the ``microcanonical''
	distribution (------), ``Cohen's energy distribution'' (--- - ---)
	and the ``product'' distribution (- - -). The thin vertical line
	marks the average binding energy of $-2.85$ a.u.}
	\label{fig:EDistr}
\end{figure}
\showlabel{fig:EDistr}

In addition to the density in phase space the energy spread
$\sqrt{\Delta E_t^2} = \sqrt{\langle E_t^2 \rangle - \langle
E_t\rangle^2}$ of the distributions turned out to be an important
criterion, too.  All three distributions have the same total energy of
$-2.85$ a.u. = $-$77.5 eV, but their energy spread is quite different,
see figure \ref{fig:EDistr}: the product distribution includes a rather
wide range of energies with a spread of $\sqrt{\Delta E_t^2} = 5.71$
a.u. = 155 eV, which is twice the total energy of the target.  Next
comes Cohen's energy distribution, which has $\sqrt{\Delta E_t^2} =
1.58$ a.u. = 43 eV. The spread of the microcanonical distribution is
determined only by the variation in the distance between the two
electrons; it has consequently the smallest value of only 1.18 a.u. =
32 eV.

In the next section we will confirm that the actual value of the total
energy of the initial distribution is of secondary interest only,
because the correct ionization potential is used to extract the cross
sections, not the (wrong) binding energy of the initial state.

\subsection{The cross section}

In the quantum mechanical formulation the fully differential cross
section is calculated from the squared overlap between the scattered
wave function $|\psi_s\rangle = \hat{S}|\psi_i\rangle$ and the given
final state $|\phi_f\rangle$. This overlap scalar product then
translates into a phase space integral of the scattered distribution
$\rho_s$ times the classical approximation of the projector onto the
final state $\mathcal{P}_f$.
\be{eq:FinalSP}
	|\langle \phi_f | \psi_s \rangle |^2 
	\quad \Rightarrow \quad
	\int\!\! dr\,dp\,\mathcal{P}_f \rho_s
\ee
According to the final state Hamiltonians for single and double
ionization, equations \eq{eq:HFinal1} and \eq{eq:HFinal2}, the free
electron wave functions are plane waves. The corresponding projector
consequently is a product of momentum delta functions, which are then
reformulated in final angles and energies.

We include the symmetrization of the electrons into the final state,
which means that in the classical approximation we have to consider all
permutations of the electrons in $\mathcal{P}_f$ when calculating the
overlap scalar product.

We see that the cross section is determined at first by the
(discretized) scalar product between wave functions, translated into
phase space, and only secondly by the acceptance region of a
macroscopic classical detector.  This is not a technical difference to
the classically derived standard CTMC, but one of interpretation: in
this quasiclassical context one should refrain from ``interpreting''
single trajectories as the paths of real electrons, they are only 
time dependent discretization points in phase space.

From equation \eq{eq:FinalSP} we easily see how to deal with excitation
cross sections in this quasiclassical approach: the final state
projector $\mathcal{P}_f$ then explicitly includes the final bound
state of the excited target electron according to the translation rule
chosen for the initial state (cf.  section \ref{sec:InitialState}).  To
calculate the excitation cross section the final values of all
trajectories are summed up with a weight according to the final excited
state's phase space density.  Special care has to be taken to ensure
that the phase space densities chosen to represent the bound orbitals
are orthogonal in order to conserve probability.

An ansatz to use final state densities modelled after the quantum wave
functions along the lines of Eichenauer \etal \cite{EIC81} and Hardie
and Olson \cite{HAR83} was recently introduced by Sattin \cite{SAT03}. 
There microcanonical distributions are added up according to a weight
function depending on the orbits' energy.

The classically derived prescriptions for bining the trajectories' 
final values, on the contrary, are based on action angle quantization 
\cite{BEC84, RAK01}.

The final state now has a fixed total energy, but our initial state,
and consequently the scattered distribution $\rho_s$, too, is not
confined to the energy shell. The overlap with the on-shell final
state would consequently cut out of the whole $\rho_s$ only those
trajectories, which initially had started on the energy shell, i.e.,
only a part of the initial distribution would actually be used. This
is both a conceptually unsatisfying situation and a problem of
normalization.

To solve this issue we switch to cross sections differential in energy
transfer, as explained for the hydrogen target in reference
\cite{GEY02}.  The energy difference $\Delta E_p$ for the projectile
electron is defined as the difference between the final and the initial
energy,
\be{}	\Delta E_p = E_p^f - E_p^i, \ee
with $E_p^f = \frac{p^2_f}{2}$ and $E_p^i = \frac{p^2_i}{2}$. For the
electrons of the helium target the electron--electron repulsion in the
initial state has to be accounted for, too, as it is released in the
ionization process. For the initially bound electrons the energy
transfer is consequently defined as, here written without the
subscripts 1 or 2:
\be{eq:DeltaETarg}
	\Delta E = E^f - E^i 
		=\frac{p_f^2}{2} - \left( \frac{p_i^2}{2} - \frac{Z}{r_i}
		+ \frac{1}{2} \frac{1}{r_{12}} \right)
\ee

All final states, which contribute to double ionization, are
characterized by three free electrons, i.e., 
\be{eq:DoubleCond}
	-\Delta E_p > IP_1 + IP_2 \quad \mbox{and} \quad
	\Delta E_1, \Delta E_2 > \frac{IP_1+IP_2}{2}\, .
\ee
$IP_1=24.6$ eV and $IP_2=54.4$ eV are the (positive) ionization
potentials for the first and the second electron, respectively: the
projectile has to supply enough energy to ionize both electrons.

For single ionization two of the three electrons are free and the 
third is still bound, i.e., we get the following conditions for the 
energy transfers, which have to be tested with electrons 1 and 2 
exchanged, too:
\be{eq:SingleCond}
	-\Delta E_p > IP_1, \quad
	\Delta E_1 > \frac{IP_1+IP_2}{2} \quad \mbox{and} \quad
	\Delta E_2 < \frac{IP_1+IP_2}{2}
\ee
These requirements may seem asymmetric at first sight, but when one
electron is ionized, the interaction energy between the target
electrons, which is the difference between $IP_1$ and $IP_2$, is
released, too, making up for the difference between $\Delta E_p$ and
$\Delta E_1 + \Delta E_2$ (see equation \eq{eq:DeltaETarg}).

Apart from this selection through the final state another constraint 
has to be considered, which originates in the M\o{}ller propagation 
scheme: the central forward propagation can only be reversed into the 
final backward leg, if those interactions, which have to be switched 
off from $H$ to $H_0^{f(1)}$ or $H_0^{f(2)}$, respectively, have 
vanished sufficiently. To turn off the nucleus--electron interaction 
an electron has to be far away from the nucleus, but it can only get 
away if it is not bound, i.e., if its actual energy is 
positive:
\be{eq:FreeFinal}
	E^f = \frac{p_f^2}{2} - \frac{Z}{r_f} > 0.
\ee

For the total cross section for single or double ionization therefore
the summation over contributing final states \eq{eq:FinalSP}
satisfying the constraint from the M\o{}ller scheme \eq{eq:FreeFinal}
amounts to counting those trajectories, for which both the energy
transfers, equations \eq{eq:SingleCond} and \eq{eq:DoubleCond}, are
correct and the actual energies \eq{eq:FreeFinal} of two or three
electrons are positive, respectively. With these sums, $N^{(1)}$
and $N^{(2)}$, the absolute cross sections are calculated in the usual
CTMC way from the maximum impact parameter $B_0$ and the total number
$N$ of propagated trajectories:
\be{eq:WQs}
	\sigma^{(1)} = \frac{\pi B_0^2}{N} N^{(1)} 
	\quad \mbox{and} \quad
	\sigma^{(2)} = \frac{\pi B_0^2}{N} N^{(2)}
\ee
The single ionization cross section also includes events, in which the 
remaining electron is excited.

The recipe for how to extract cross sections is transferred into the
classical form independently of the initial state description. This is
the reason why we do not refer to the smeared binding energy $E_t$ of
the initial distribution, but to the experimentally observed correct
ionization potentials $IP_1$ and $IP_2$. Any initial state can be
used, independent of its energy; one might even fancy two electrons
without a nucleus localized in space according to the quantum
densities --- a setup with a positive total energy. But an electron
would be considered ionized only when enough \emph{additional} energy
has been transferred onto it.

\section{Results}

Up to now we have repeated and detailed how the quasiclassical
procedure has to be set up to calculate electron impact ionization on
atomic helium. That the propagation scheme actually refocuses the
autoionizing helium target has been shown in a recent letter
\cite{GEY03}. We now present and discuss the total cross sections for
single and double ionization and then confirm our findings through
differential cross sections.

\subsection{Total cross sections}

\begin{figure}[t]
  \centerline{\epsfbox{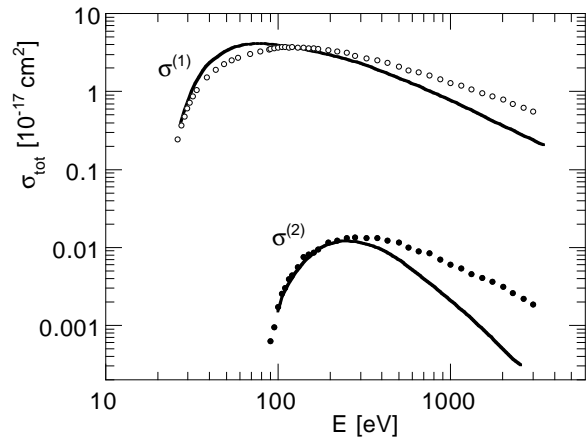}}
  \caption{Total cross sections $\sigma^{(1)}$ for single and 
  $\sigma^{(2)}$ for double ionization: Comparison of our results 
  (solid lines), calculated with the ``product'' distribution (see 
  text), with the experimental data of Shah \etal \cite{SHA88} (open 
  and filled circles).}
  \label{fig:Stot}
\end{figure}

With the three initial distributions explained above the total
ionization cross sections $\sigma^{(1)}$ for single ionization and
$\sigma^{(2)}$ for double ionization are calculated. The results are
compared to measurements by Shah \etal \cite{SHA88} in figure
\ref{fig:Stot}. The differences between the initial distributions are
only small both for $\sigma^{(1)}$ and $\sigma^{(2)}$: all three
distributions lead to the same shape of the cross sections while the
absolute value differs in a range of about $\pm$15\%. Therefore figure
\ref{fig:Stot} only shows the results from the product distribution.
The result from the microcanonical distribution is slightly higher,
while Cohen's energy distribution leads to a slightly smaller total
cross section. A similar behavior had occurred already for the total
cross sections on hydrogen, calculated in \cite{GEY02}.

\epsfxsize=8cm
\epsfysize=6cm

The single ionization cross section $\sigma^{(1)}$ agrees with the
measurement on the level expected from a classical trajectory
calculation: the overall coincidence is reasonable, but the absolute
value at the maximum is slightly too high, the maximum itself is
shifted to lower energies by a factor of about two and the high energy
behavior follows the classical $1/E$ decay \cite{THO12} instead of the
correct Bethe--Born limit of $\ln (E)/E$ \cite{INO71}.

This cross section had been calculated previously by Schultz
\emph{etal}, both in the independent electron model and with the nCTMC
procedure \cite{SCH92}.  Our calculation with the microcanonical
distribution reproduces the nCTMC result, which is closer to the
experiment than the IEM model.  This is not surprising, as the nCTMC
initial state can be viewed as a symmetric subset of our microcanonical
description.  The additional stabilization potential used there
prevents autoionization before the scattering event and then
essentially vanishes, once one of the target electrons is ionized.

Before we actually look at our results for the double ionization cross
section $\sigma^{(2)}$, we want to contemplate, which high energy
behavior can be expected from a classical description: again, as with
$\sigma^{(1)}$, quantum effects like tunneling are not accounted for in
the classical treatment.  Additionally the so called shake--off
process, which is dominant for double ionization at very high energies
\cite{MCG82}, can not be modelled classically: When in the classical
description one of the electrons is removed then there is no
quantization condition as in quantum mechanics, which ``forces'' this
electron to change its orbit to one of the allowed ones --- either to
one of the discrete bound levels or to the continuum.  Classically any
orbit at any binding energy is allowed, so there is no need for a
transition; and consequently no shake--off process.  In the classical
treatment even an opposite effect occurs: when one of the target
electrons is removed, the remaining ion is actually ``stabilized'' in
its current (excited) state, because with the other target electron the
cause for autoionization is removed.

Though the shake--off process does not exist in the classical
description, there are still two important processes left, which lead
to double ionization: in the so called ``Two--Step 1'' (TS1) process
the projectile ionizes one of the target electrons, which in turn
ionizes the other one in a subsequent $(e,2e)$ reaction. The
probability for the second ionization only depends on the energy
spectrum of the first electron and is therefore nearly independent of
the projectile energy. In the classical description the TS1 process
consequently should decrease roughly proportional to the single
ionization's $1/E$ law. The other process, the ``Two--Step 2'' (TS2)
process, consists of two independent encounters between the projectile
and the two target electrons. Each collision's probability falls off
as $1/E$, so the TS2 decays proportional to $1/E^2$. Therefore it can
be deducted from the high energy behavior of $\sigma^{(2)}$, if the
TS1 or the TS2 process is dominant in the classical description.

Looking at the results for the double ionization cross section
$\sigma^{(2)}$ in figure \ref{fig:Stot} we see that for high energies
it decreases much faster than $\sigma^{(1)}$ --- in fact it follows a
$1/E^2$ behavior for $E>600$ eV --- while the experimental results
confirm the constant asymptotic ratio of $\sigma^{(2)}/\sigma^{(1)}$
derived from quantal calculations \cite{MCG82}.

We therefore may conclude, that in classical trajectory calculations
the double ionization at very high energies mainly happens through the
``Two--Step 2'' (TS2) mechanism, whereas the TS1 seems to be missing. 
To verify this hypothesis, which contradicts the experimental
evidence \cite{DOR02}, we have to look at differential cross
sections.

\epsfxsize=8cm
\epsfysize=6cm
\begin{figure}[t]
  \centerline{\epsfbox{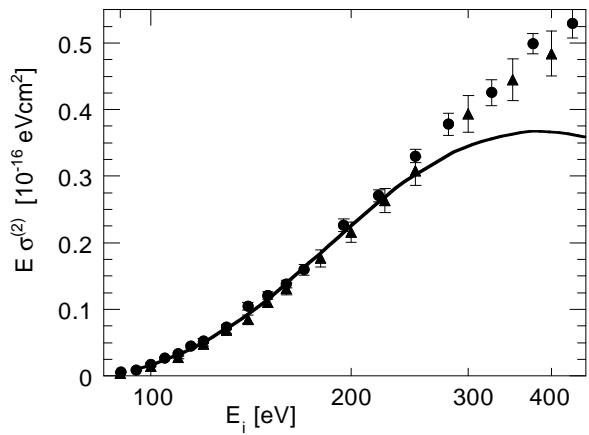}}
  \caption{Bethe like plot of the double ionization cross section
  $\sigma^{(2)}$: comparison of our quasiclassical calculation with
  the product distribution (solid line) to the experimental results of
  Shah \etal \cite{SHA88} (circles) and of Rejoub \etal \cite{REJ02}
  (triangles). For better comparison with figure \ref{fig:Stot} the
  impact energy is not scaled with the ionization threshold.}
  \label{fig:Bethe}
\end{figure}

For energies below the maximum of the cross section, on the other
hand, the experimental results of $\sigma^{(2)}$ are much better
reproduced than for $\sigma^{(1)}$.

The correspondence between the quasiclassical results and the
experimental data at low energies can be seen more pronounced, if a
Bethe plot like presentation is chosen: figure \ref{fig:Bethe} plots
the product of the total energy and $\sigma^{(2)}$ against the impact
energy, our results with the product distribution as well as
experimental data from Shah \etal \cite{SHA88} and Rejoub \etal
\cite{REJ02}.  The cross section is in nearly perfect agreement with
the measurements from low impact energies up to about 250 eV. For
impact energies above 250 eV our calculation falls back against the
experimental results.  Against common experience and the usual
interpretation of the correspondence principle high energies in
electron impact ionization obviously do not mean that the system
becomes more ``classical'' --- on the contrary, the description becomes
worse and worse, the higher the energies become.  Further classical
calculations should therefore focus on the low energy regime, where the
correlated dynamics of all four particles becomes more and more complex
and even chaotic.  Therefore it is not possible to make a simple
prediction about the ionization mechanism at intermediate and low
energies.  It is only right above the threshold where a highly
symmetric final state configuration and a power law behavior is
predicted by a Wannier type argument \cite{PAT98,BAR03}.

With these arguments it is understandable that in the classical
calculation $\sigma^{(2)}$ is reproduced much better at low energies
than $\sigma^{(1)}$: due to the higher dimensionality of the phase
space there exist more classical reaction paths for double than for
single ionization, which renders quantum effects less important than
when only one electron is removed.

If we attribute the difference between our results and the experiment
to the missing TS1 process, as explained above, then we can estimate
that the probability for the sequential TS1 is negligible below about
250 eV. For a TS1 event to be possible at 250 eV the projectile has to
transfer about one third of its energy onto one of the target
electrons, which is already a rather ``violent'' collision. For higher
impact energies the TS1 process then becomes increasingly important
and we can estimate from the difference between our calculated
$\sigma^{(2)}$ and the experiment that at $E_i \approx 500$ eV both
processes contribute with equal weight to double ionization. With
fully differential cross sections a ``great importance of second-- or
higher--order effects'' at 600 eV has been observed by Lahmam--Bennani
\etal \cite{LAH02}.

For double ionization there is, for obvious reasons, no CTMC
comparison available yet. Our result is the first classical trajectory
calculation for double ionization with two active target electrons
without additional stabilization potentials or the need to ``switch
off'' the interaction between the target electrons.

It should be emphasized, that the absolute cross sections for both
single and double ionization can be calculated from the same run of
the program. For both cases the same initial distribution is used and
both target electrons are explicitly included. In principle excitation
could be included in the analysis, too, though here we confine
ourselves to ionization alone. This is parallel to the experiment,
where single and double ionization and excitation all occur
concurrently.

It was not clear a priori, though, if this would really work. For
single ionization, e.g., the second electron, which is not ionized,
might not act as an independent spectator and could disturb the single
ionization process. 

In an IEM treatment it is for methodological reasons not possible to
calculate all processes simultaneously: there the electron--electron
repulsion in the target is neglected and therefore only one of the two
thresholds for single \emph{or} for double ionization can be described
correctly. The same difficulty arises in the nCTMC treatment due to
the restricted symmetric geometry of the initial distribution.

In an actual calculation it is more efficient, though, to run the
calculation twice with two different maximal impact parameters: a
bigger one for $\sigma^{(1)}$ to cover the whole target area and a
smaller one to achieve enough statistics for the much rarer double
ionization, where both target electrons have to be located around the
path of the projectile. Typical maximal impact parameters are $B_{0}
\approx 30$ a.u. for single and $B_0 \leq 3$ a.u. for double
ionization. Then at each impact energy about $10^5$ trajectories are
necessary for sufficiently small statistical errors of the total cross
section.

The propagation of the electrons' trajectories is performed by a
symplectic integrator, which separates the attractive regularized
electron--nucleus interaction from the electrons' mutual repulsion (for
details please see reference \cite{GEY02}).  This algorithm is
extremely stable, as it even allows the electrons to fall ``into'' the
nucleus.  To calculate differential cross sections huge numbers of
trajectories have to be run and the number of faulty trajectories has
to be much smaller than the few trajectories that contribute to a given
cross section.  Typical fractions of trajectories which had to be
discarded are on the order of one per every ten million trajectories,
which is smaller than the yield of double ionization events by at least
three orders of magnitude.

\subsection{Differential cross sections at high energies: $E_i=2$  keV}

The total double ionization cross section at high energies falls off 
as $1/E^2$, which is much faster than the experimental data. This had 
lead us to conclude that in this regime double ionization essentially 
occurs through independent collisions between the projectile and each 
of the two target electrons. But now we need to verify this 
conjecture, because the difference between our calculation and the 
experiment might also be caused by the classical approximation 
itself, i.e., that the occurring processes are modelled correctly, but 
only the overall probability is too small due to the description of 
the electrons as classical point particles.

We therefore ran calculations at an impact energy of 2 keV, which is
well into the high energy regime. At this energy single and double
differential cross sections were measured by Dorn \etal \cite{DOR99}.
The numbers of trajectories calculated with the three initial state
descriptions and the resulting double ionization events are given in
table \ref{tbl:Numbers2keV}. To increase the number of double
ionization events the impact parameter of the projectile was sampled
linearly between 0 and $B_0$; in the analysis each trajectory then
is weighted with an additional factor of $\frac{b}{2B_0}$.
Therefore the total double ionization cross section can not be
calculated from the values given in table \ref{tbl:Numbers2keV} alone
but only from the saved trajectory data.

\begin{table}[tbp]
	\centering
	\begin{tabular}{l|ccc}
		& $N_t$ & $N_2$ & $B_0$  \\
		\hline
		Cohen & 310 Mio. & 33447 & 1.2 a.u.  \\
		product & 39 Mio. & 7617 & 1.5 a.u.  \\
		microcan. & 50.6 Mio. & 19740 & 1.2 a.u.  \\
	\end{tabular}
	\caption{Number of trajectories $N_t$ and double ionization events
	$N_2$ for the three initial state distributions at $E=2$ keV
	together with the maximal impact parameter $B_0$ (see text).}
	\label{tbl:Numbers2keV}
\end{table}
\showlabel{tbl:Numbers2keV}

The two ``classical'' ionization processes, the TS1 and the TS2, lead 
to different signatures in the angular and energetic distribution of 
the ejected electrons. They are already discernible in partially 
differential cross sections: at this high impact energy the energy 
loss of the projectile is much smaller than its initial energy, the 
final channel therefore consists of one fast and two slow electrons. 
These two slow electrons repel each other on their way out; ejection 
into the same half sphere is consequently suppressed. Due to the two 
independent collisions this is the main angular signature of the TS2 
process. In the TS1 process, on the other hand, the momenta of the two 
slow electrons are correlated because of the second $(e, 2e)$ process. 
This elastic collision between two particles of identical mass leads to 
a peak in the relative angle between the slow electrons in the range 
between about 90 and 110 degrees, depending on the relative energy 
partitioning and the momentum transferred.

\epsfxsize=8cm
\epsfysize=6cm
\begin{figure}[t]
  \centerline{\epsfbox{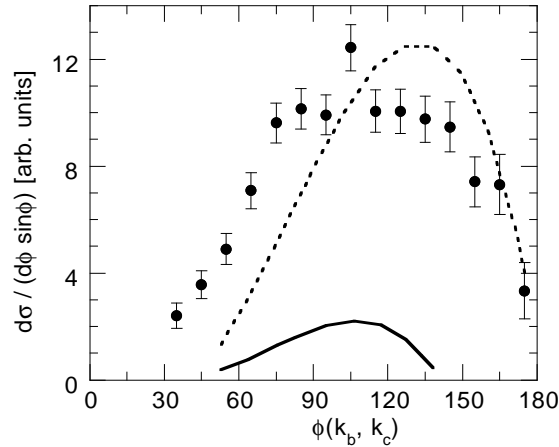}}
  \caption{Angle $\Phi$ between the two slow ejected electrons at
  double ionization with $E_{in} = 2$ keV integrated over all energies
  $E_b+E_c <35$ eV: comparison of our results (full circles) with
  polynomial fits of the experimental results of Dorn \etal
  \cite{DOR99} ($E_b+E_c < 20$ eV) for small recoil ion momentum (solid
  line) and for small momentum transfer (broken line).}
  \label{fig:theta23}
\end{figure}

This angle between the two slow electrons is plotted in figure
\ref{fig:theta23} for an impact energy of 2~keV: our results are
compared to polynomial fits of the measurements of Dorn \etal
\cite{DOR99}.  The relative angle is integrated over all pairs of slow
electrons, which have a sum energy of $E_b + E_c < 35$ eV in the
calculation and $E_b + E_c < 20$ eV in the experiment.  The
experimental data are given for two different kinematical regimes,
discerned by the momentum transfer of the projectile $q$ and the recoil
ion's momentum.  Both the experimental curves show a distinct peak,
whose position varies with the momentum transfer.  Our results, on the
other hand, show a broad distribution of angles, whose only feature is
the suppression of small relative angles due to the electrons' mutual
repulsion.  The form of the angular distribution is independent of both
the momentum transfer and the maximum sum energy, only the number of
contributing events varies.  The angular distribution is essentially
independent of the relative energy partitioning $\epsilon =
\frac{E_b}{E_b+E_c}$, too.

This behavior is consistent with the hypothesis that at high energies 
in the classical descriptions the TS2 process is dominant, contrary to 
the experimental evidence.

\epsfxsize=12cm
\epsfysize=6cm
\begin{figure}[t]
  \centerline{\epsfbox{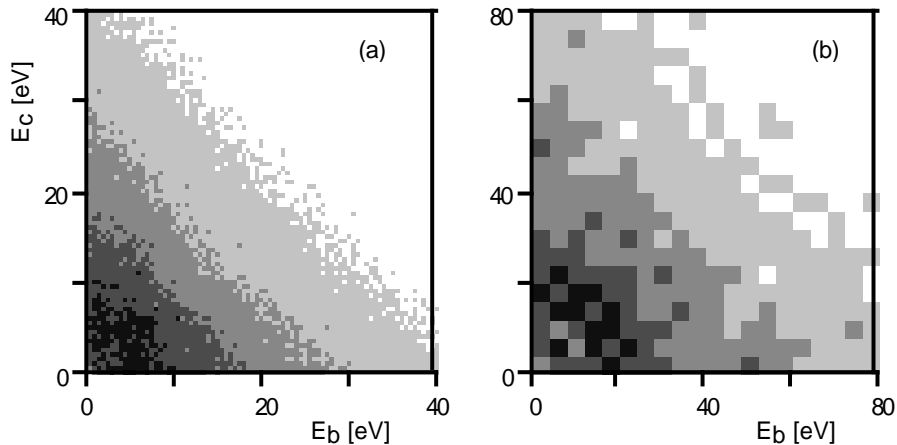}}
  \caption{Joint distribution of the energies $E_b$ and $E_c$ of the
  two slow electrons for $E_{in} = 2$ keV: comparison of the
  experimental result of Dorn \etal \cite{DOR02b} to our calculation
  with Cohen's distribution.  Note that the energy scales of the two
  plots differ by a factor of two.}
  \label{fig:EbEc-Dorn-Cohen}
\end{figure}
\showlabel{fig:EbEc-Dorn-Cohen}

The two considered processes, the TS1 and the TS2, can be discerned,
too, by comparing the joint distribution of the slow electrons' final
energies: at these high impact energies the energy transfer is much
smaller than $E_i$, so that on the average in the two collisions of
the TS2 twice as much energy is transferred onto the target electrons
than in the one collision of the TS1. In the TS1 this energy is then
shared between the two target electrons. The energy distribution
of electrons ionized in a TS2 process should therefore be about twice
as wide as the one of electrons from the TS1 process.

The distribution of energies of the two slow electrons is plotted in
figure \ref{fig:EbEc-Dorn-Cohen}.  Panel (a) shows the experimental
data of Dorn \etal \cite{DOR02b}, whereas our quasiclassical result
with Cohen's initial distribution is shown in (b).  At first glance
both results seem to be similar, but it is important to note that in
(b) the scale of the energy axes is twice as wide as in (a).

The dip in the experimental cross section where both energies are
small is due to the time resolution of the spectrometer: events, in
which the longitudinal momentum difference between the slow electrons
is less than $|p_b - p_c| < 0.3$ a.u. could not be resolved in this
experiment. When both energies are small, then the relative momentum
is also small. This restriction is included in our result, too.

\epsfxsize=12cm
\epsfysize=6cm
\begin{figure}[t]
  \centerline{\epsfbox{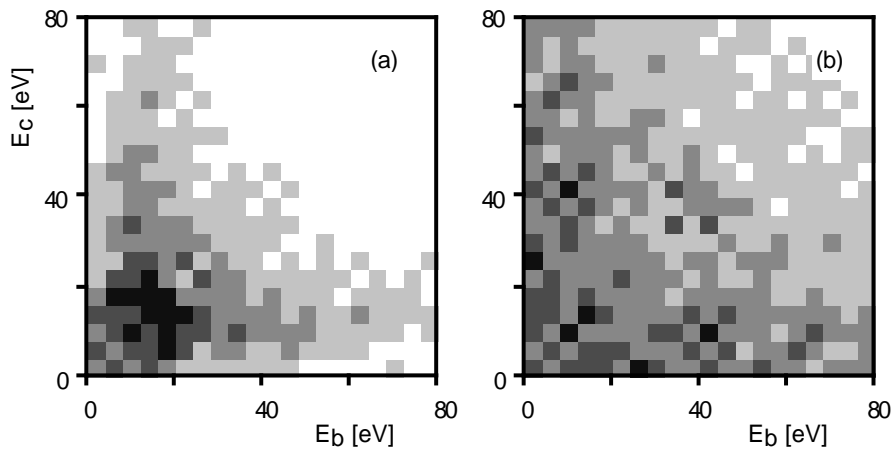}}
  \caption{Same as figure \ref{fig:EbEc-Dorn-Cohen}: results from the
  calculations with (a) the microcanonical initial distribution and (b)
  with the product target description.}
  \label{fig:EbEc-micro-prod}
\end{figure}
\showlabel{fig:EbEc-micro-prod}

Figure \ref{fig:EbEc-micro-prod} shows the same cross section
calculated with the microcanonical distribution (a) and with the
product initial state description (b).  The effect of the different
initial state descriptions can clearly be seen: with the microcanonical
distribution the very asymmetric events are missing, i.e., when one of
the ejected electrons only has a small energy.  The explanation is
connected to the form of the energy spread of the initial distribution,
see figure \ref{fig:EDistr}: both electrons are sampled independently
at the same fixed one--electron energy.  Consequently the spread of the
total energy is determined solely by their interaction --- but as their
distance is limited there is a lower limit for this contribution, and
therefore for the total energy of the target initial conditions.  There
is no upper limit for the electron--electron repulsion, therefore, to
achieve the correct average binding energy, most of the initial values
must have a binding energy of less than the average to compensate for
the relatively few events, where the interaction energy is high.  These
initial conditions with energies below the average are harder to ionize
within our scheme, which requires the electrons to be actually free in
the end.  As the final energies increase this threshold blocks less and
less events that have the correct energy transfer and then the cross
sections resemble that from Cohen's distribution (figure
\ref{fig:EbEc-Dorn-Cohen}(b)).

This again shows that an initial distribution which is constructed
from a classical recipe may have unphysical properties --- here the
hard limits in the radial and the energy distributions --- and lead to
problems, when used in a quantum mechanically derived context.

With the product distribution, figure \ref{fig:EbEc-micro-prod}, the
energies of the two slow electrons are spread out to even higher
values compared to Cohen's distribution. This is the consequence of
the very broad energy spread of the initial state, which is about
twice as wide as the distribution of final energies resulting from the
ionization process. But it is not our method which breaks down. The
energy of each trajectory is still well defined. This very wide
initial energy spread means that there are many ``extreme'' initial
values, i.e., trajectories, in which the electrons are started with
high momenta at large distances from the nucleus, with low momenta at
small distances or with a very small interelectronic distance.
Trajectories started there are far more likely to end ``outside'' the
main features of the cross section and therefor blur the result.

Summarizing, one can say that the evidence presented --- the high
energy behavior of $\sigma^{(2)}$, the distribution of the
inter--electron angle and the energies of the slow electrons ---
confirms that at high energies a (quasi) classical treatment of double
ionization can not describe the important processes.  In general the
coupling between the target electrons is grossly underrepresented.  The
reason for this behavior originates in the description of the electrons
as point particles: when the first electron is not kicked exactly into
the direction of the other target electron, the second stage of the TS1
process can not take place.  Obviously, in most cases the first
electron is ejected so that the other electron does not ``notice'' it.
Or in other words: for the first electron to ionize the other, it has
to be kicked into a very narrow cone around the vector connecting the
two electrons' positions.  In the quantum mechanical treatment both
electrons are located spherically symmetric around the nucleus; the
first ionized electron always has to pass ``through'' the other
electron.  Therefore the subsequent ionization can take place
independent of the direction in which the first electron leaves the
atom.

This shortcoming of the classical approximation is independent of the
initial state's description, though the results show that it is
nevertheless important to choose the best possible distribution: an
inappropriate initial distribution can only make the results worse.

It is interesting to note that recently photo double ionization has
been described amazingly well in a mixed quantum classical picture
\cite{SCH02}.  In the classical part, which models the sequential
``TS1'' process the electron that absorbs the photon starts ``on'' the
nucleus.  From this highly symmetric starting point it always has to
pass through the probability distribution of the other electron and the
chance for their encounter is independent of the initial direction of
the first electron.

\subsection{Differential cross sections at $E_i=250$ eV}

For low impact energies up to about 250--300 eV the agreement between
the measured and the calculated total double ionization cross section
is remarkably good, see figure \ref{fig:Bethe}, suggesting that in
this energy regime the quasi classical treatment should be able to
describe those dynamics, which are actually taking place.
Unfortunately there are no measurements of electron impact ionization
of helium available yet in this energy regime, though first results of
kinematically complete experiments at 500 eV are just being reported
\cite{DOR03} and results from even lower impact energies will surely
follow. We decided not to perform calculations at $E_{in}=500$ eV at
this point, because of what we learned above from the results at
$E_{in}=2$ keV we expect them to be at most partly correct. So we can
not compare our results to an experiment yet, but it will nevertheless
be instructive to look at the available quasiclassical results and
predictions.

The numbers of trajectories calculated and the resulting double
ionization events for the three initial distributions are summarized
in table \ref{tbl:Numbers250eV}. The impact parameter was sampled
linearly again as for 2 keV.

\begin{table}[tbp]
	\centering
	\begin{tabular}{l|ccc}
		 & $N_t$ & $N_2$ & $B_0$  \\
		\hline
		Cohen & 64.3 Mio. & 83581 & 1.5 a.u.  \\
		product & 130 Mio. & 35785 & 3 a.u.  \\
		microcan. & 34.8 Mio. & 49277 & 3 a.u.  \\
	\end{tabular}
	\caption{Number of trajectories $N_t$ and double ionization events
	$N_2$ at $E_{in} = 250$ eV for the three initial state
	distributions together with the maximal impact parameter $B_0$ (see
	text).}
	\label{tbl:Numbers250eV}
\end{table}
\showlabel{tbl:Numbers250eV}

The different spatial form of the microcanonical distribution, which
was explained in section \ref{sec:InitialState}, is reflected in these
number, too: to achieve a comparable number of double ionization
events $N_2$ with the product distribution about four times as many
trajectories had to be run. The resulting total $\sigma^{(2)}$ is
nearly the same, though. This means that with the microcanonical
distribution double ionization on average takes place at an impact
parameter, which is about four times larger --- the target electrons
are found at larger radii.

Now, at $E_{in}=250$ eV, the projectile electron is only about twice
as fast as the classical target electrons; the dynamics is much slower
than at 2~keV and it can be expected that the distinction between the
fast projectile and the slow ionized target electrons vanishes.

\epsfxsize=8cm
\epsfysize=5cm
\begin{figure}[t]
	\centerline{\epsfbox{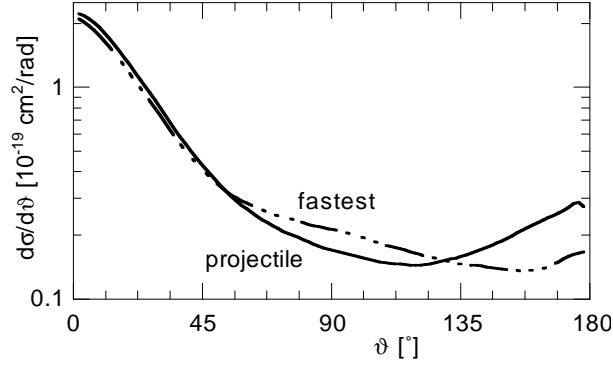}}
	\caption{Angle differential cross $\frac{d\sigma}{d\theta}$ at
	$E_{in}=250$ eV for the fastest final electron (broken line) and
	the former projectile electron (solid line) after a double
	ionization event.}
	\label{fig:thetafast}
\end{figure}
This effect can already be seen when the angle differential cross
sections $\frac{d\sigma}{d\theta}$ of the fastest final electron and of
the former projectile after a double ionization event are compared, see
figure \ref{fig:thetafast}.  At $E_{in}=2$~keV the two cross sections
are indiscernible (not shown here), but now they differ by about 10~\%
in the forward direction and even more for angles larger than
$60^\circ$.  The projectile is still emitted preferentially into the
forward direction, but in one of about every ten double ionization
events it transfers so much of its energy onto the target electrons
that one of these becomes the fastest.  These fast ionized target
electrons then show up in figure \ref{fig:thetafast} in the direction
perpendicular to the incident projectile, i.e, at angles between
$60^\circ$ and $120^\circ$.  When the projectile is scattered
backwards, i.e., $\theta > 135^\circ$, it had to reverse its momentum.
This big energy transfer makes that in the backward direction the
projectile is rarely the fastest electron.

\epsfxsize=8cm
\epsfysize=5cm
\begin{figure}[t]
	\centerline{\epsfbox{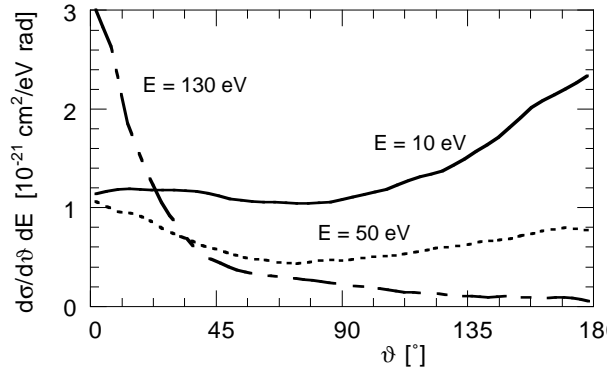}}
	\caption{Doubly differential cross section
	$\frac{d\sigma}{d\theta\, dE}$ for double ionization at
	$E_{in}=250$ eV for the fixed final electron energies of 10 eV
	(------), 50 eV (- - -) and 130 eV (--- - ---).  The calculation
	was performed with Cohen's distribution.}
	\label{fig:dsdtheta}
\end{figure}

In an experiment the distinction between the former projectile and the
target electrons is not possible. The angular distribution of the
electrons is usually measured for a fixed final energy. Of course,
this type of cross section can be extracted from our calculations,
too, see figure \ref{fig:dsdtheta}: it plots the absolute doubly
differential cross section $\frac{d\sigma}{d\theta\, dE}$ for double
ionization, integrated over all three electrons, for the fixed final
energies of 10 eV, 50 eV and 130 eV, calculated with Cohen's initial
distribution. Again, as already in figure \ref{fig:thetafast}, the
fast electrons (130 eV) are emitted preferentially into the forward
direction, while the slowest electrons (10 eV) are ejected into the
rear half sphere. The crossover between the emission in forward and in
backward direction takes place at an electron energy of 50 to 60 eV,
which is about a third of the total energy of the system.

\epsfxsize=6cm
\epsfysize=6cm
\begin{figure}[t]
	\centerline{\epsfbox{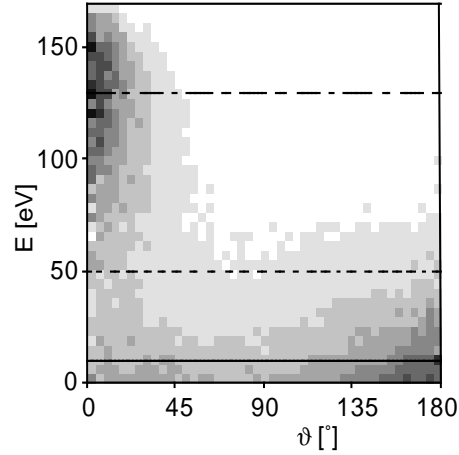}}
	\caption{Doubly differential cross section
	$\frac{d\sigma}{d\theta\, dE}$ at $E_{in} = 250$ eV, integrated
	over all electrons.  The greyscale is linear between 0 and the
	maximum value in arbitrary units.  The three horizontal lines
	correspond to the cross sections at the fixed energies plotted in
	figure \ref{fig:dsdtheta}.}
	\label{fig:Etheta}
\end{figure}

An overview over the behavior of $\frac{d\sigma}{d\theta\, dE}$ is 
given in figure \ref{fig:Etheta}: it plots this cross section as a 
function of both the angle $\theta$ and the final one electron 
energy $E$. The dominant structures of this cross section are the 
maximum for fast electrons with an energy around 130 eV and small 
angles up to about $20^\circ$ and the broad angular distribution of 
the slow electrons with an emphasis on the backward direction. 
As said before, the crossover between the two angular patterns takes
place around 50 eV.

Both figure \ref{fig:dsdtheta} and \ref{fig:Etheta} are not very
sensitive to which initial distribution has been used; the differences
are the same as those already observed at the slow electrons' joint
energy distributions, see figures \ref{fig:EbEc-Dorn-Cohen}(b) and
\ref{fig:EbEc-micro-prod}: with the microcanonical distribution the
events with very small final energies are missing again and with the
product distribution both the distribution of the angles and the
energies is broadened.

\epsfxsize=12cm
\epsfysize=6cm
\begin{figure}[t]
	\centerline{\epsfbox{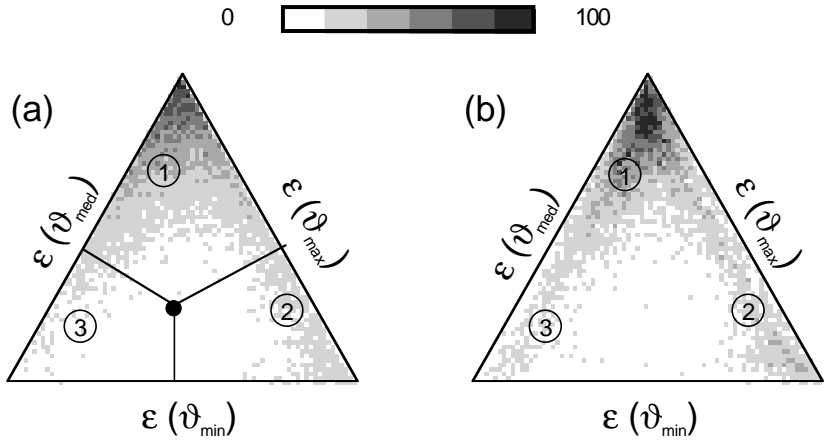}}
	\caption{Dalitz plot of the three electrons' final energies,
	calculated with Cohen's distribution (a) and with the
	microcanonical initial distribution (b). The greyscale is linear
	between 0 and the maximal value for each plot. The different
	energy partitions, which correspond to the numbers inside the
	triangles, are explained in the text.}
	\label{fig:ETriangles}
\end{figure}

The cross sections at $E_{in}=250$~eV shown until now were extracted 
from the double ionization events, but they did not test for any 
correlation between the outgoing electrons. 

At 2~keV the projectile is much faster than the ionized target
electrons; the correlation between the electrons affects
mainly the ``subsystem'' of the two slow electrons. The presence of the
fast projectile can conveniently be reduced to a momentum and energy
transfer onto the otherwise independent target system. But now, at
$E_{in}=250$~eV, where the dynamics is much slower and the three final
energies are closer together this separation is at best questionable.
We therefore need a way to plot the final energies and angles, which
is sensitive to correlations between the three particles.

One possibility is the so called Dalitz plot \cite{DAL53}: it places
the relative final energies $\epsilon_i = \frac{E_i}{E_a+E_b+E_c}$,
with $i=a,b,c$, of the three electrons inside an equilateral triangle,
cf.  figure \ref{fig:ETriangles}.  As
$\epsilon_a+\epsilon_b+\epsilon_c=1$ each combination of final energies
is mapped unambiguously onto one point inside the triangle, for which
the distances from the three sides have the same ratios with each other
as the relative energies.  This representation of the energies of three
correlated particles was first used in high energy physics, but has
recently been adapted in atomic physics, too, e.g., for the
fragmentation of $H_3^+$ \cite{WIE97} or for triple ionization by fast
heavy ion impact \cite{SCH00}.  We follow this work of Schulz \etal and
sort the final relative energies $\epsilon_i$ according to the
corresponding scattering angle $\theta_i$: the relative energy of the
electron with the smallest $\theta = \theta_{min}$ is measured from the
base of the triangle, the $\epsilon$ for the biggest $\theta =
\theta_{max}$ from the right side and the intermediate
$\epsilon(\theta_{med})$ from the left side.

Figure \ref{fig:ETriangles} shows the resulting distributions, when
Cohen's distribution is used as the target initial state in panel (a)
and with the microcanonical distribution in (b). Their overall
structure is similar: the maximum is at the top of the triangles,
which means that the highest (relative) energy mainly goes with the
smallest scattering angle. This is consistent with the cross section 
of figure \ref{fig:Etheta}.

With Cohen's distribution the plot is asymmetric with respect to
$\epsilon(\theta_{med})$ and $\epsilon(\theta_{max})$: when
$\epsilon(\theta_{min})$ is big, i.e., the fastest electron is
scattered into the forward direction, then the energy of the electron
with the biggest angle is higher than that of the electron with the
intermediate $\theta_{med}$. This region is marked with ``1'' in
figure \ref{fig:ETriangles}(a). And then there is a whole band of
events near the right side of the triangle, labelled with ``2'', where
$\epsilon(\theta_{max})$ is very small and $\epsilon(\theta_{med})$ is
the biggest. This corresponds to the central region of scattering
angles in figure \ref{fig:thetafast}: the projectile, which is not the
fastest electron any more, is emitted under small angles and one of
the target electrons got a large transfer of energy from the
projectile.

With the microcanonical distribution, figure \ref{fig:ETriangles}(b),
it can be seen again that events for small $\epsilon_i$ are missing. 
The plot is more symmetric, too.  In region ``1'' both
$\epsilon(\theta_{med})$ and $\epsilon(\theta_{max})$ are about equal,
the correlation between the scattering angle and the final energy of
the two slow electrons is nearly gone.  Region ``2'' is similar to
panel (a), though: this could already be expected from the angular
distribution of the fastest and the projectile electron --- which was
not shown for the microcanonical distribution.  This symmetry between
the two slow electrons continues from high values of
$\epsilon(\theta_{min})$ to low values: region ``3'', which is
essentially empty for Cohen's distribution, is nearly as important as
``2'': this, too, shows that the correlation between final angles and
energies is much less pronounced for the ``constructed'' microcanonical
distribution than for the more quantal distribution according to
Cohen.

The corresponding plot from the product distribution again looks like 
a blurred version of the result from Cohen's distribution: it is not 
shown here.

From figure \ref{fig:ETriangles}(a) two modes can be inferred, in
which the three electrons leave the nucleus: in region ``1'' the fastest
electron is emitted into the smallest angle, then comes the slowest
electron and the electron with the intermediate energy leaves with the
biggest angle; the slowest electron is between the faster ones.  In
region ``2'' this pattern is reversed: the fastest electron is between
the two slower ones, but again with the bias that high final energies
prefer small scattering angles.

The analogous pattern is observed when the relative angles between the
electrons' momenta in their center of mass system are plotted in a
Dalitz plot.  Then, one finds that the electron that is scattered
into the most forward direction has preferentially the highest
energy, followed by the one scattered into the most backward
direction.  The electron inbetween gets the smallest amount of energy. 
These plots are not shown here, because they provide no more
information than the ones of figure \ref{fig:ETriangles}.

Summarizing one can say that at this rather low impact energy of 250
eV, where the total energy of the whole system is just about twice as
large as the target's binding energy, the double ionization dynamics
is much more complex than at high energies.  There the projectile
is very fast and can be reduced to a sudden perturbation, whereas it
here is --- classically speaking --- only twice
as fast as the target electrons. Now after the reaction the fastest electron is not 
neccessarily the former projectile any more, as we have seen from the 
angle differential one electron cross section. This distinction is 
of course only possible in our classical treatment, but not in the 
experiment.

The doubly differential one electron cross section indicates two
emission patterns: fast electrons essentially show up in a cone of about
$20^\circ$ around the forward direction, while the slow electrons are
emitted into the rear half sphere.  The crossover between these two regimes
takes place at electron energies of 50 to 60 eV, at about a third of
the total energy.

Due to the low yield of double ionization trajectories and the high
dimensionality of the three--electron final state phase space we could
not extract fully differential cross sections from these calculations,
but correcllation effects between the electrons' final angles and
energies can already be observed: high energies go with small
angles, as already shown, and the scattering angle of the slowest
electron is between the two faster ones --- or the other
way around: the fastest electron emerges between the two slow ones.

This cross sections again emphasizes that also at this low impact
energy it is important to use an initial state distribution which
reproduces the wave function as closely as possible, i.e., one which is
derived from the quantum wave function.  We see differences in the
quality of the obtained cross sections between Cohen's distribution and
the product description, but the general results are the same, whereas
the microcanonical distribution, which is constructed from a classical
background, leads to artifacts and gives different results when it
comes to the correlation between the electrons.

\section{Summary}

In this paper we demonstrated that our quasiclassical ansatz allows to
calculate meaningful total and differential ionization cross sections
on helium targets by means of a classical trajectory method. To this
end we first explained how the algorithm \cite{GEY02}, which initially
had been developed with the hydrogen target, has to be adapted for
many electron targets. It is especially important not to neglect the
initial backward propagation that arises from translating the
scattering operator from the quantum M\o{}ller formulation to its
classical form. The resulting dynamical stabilization of the
autoionizing classical many electron target had been demonstrated
already in a recent letter \cite{GEY03}, together with the total cross
sections for single \emph{and} double ionization.

In this paper we took a closer look at the calculated total double
ionization cross section and concluded from its $1/E^2$ high energy
behavior that the dominant process for double ionization consists of
two independent encounters between the projectile and the two target
electrons (the ``TS2'' process).  This conjecture was then confirmed by
comparing our differential cross sections to experimental results at 2
keV impact energy.  Both the relative angle between the two slow
electrons and their joint energy distribution show no signatures of the
sequential ``TS1'' process, but the cross sections are easily explained
if only the ``TS2'' process takes place.

For impact energies below about 300 eV the total cross section
$\sigma^{(2)}$ reproduces the experiment amazingly well.  Therefore
classical calculations should focus on this low energy regime, where it
is hard to perform quantum calculations.  Though there are no
experiments available yet, we presented single and double differential
cross sections at 250 eV.

In a classical calculation it is possible to label the electrons.  With
the calculations at 250 eV we showed that at low impact energies the
identification of the fastest electron with the projectile starts to
break down.  Combining the three electrons' relative energies into one
plot we could identify a pattern, in which the electrons' energies are
``interleaved'' with their scattering angle out of the forward
direction: when the electrons are sorted according to their scattering
angle then either the electron with the highest or with the lowest
final energy is the middle one, but it is relatively unlikely that,
e.g., the electrons' energies decrease monotonically with increasing
angle.  Of course, we are anxious to see, whether the experiment will
confirm our predictions --- or not.

When performing the calculations on hydrogen it had been important to
choose a phase space description for the target atom that best
describes those features which are important in the corresponding
energy range.  We therefore ran the calculations on helium with three
different initial distributions, too: two derived from the quantum wave
function and one built from a classical approach.  The two quantum
derived distributions only differ in the quality of the resulting cross
sections, whereas the classically derived one leads to artifacts, which
clearly stem from its conceptual limitations.  Consequently it is
important to choose --- within the constraints of feasibility and
numerical effort --- the best initial phase space distribution possible
for the target: the classical calculation itself is a rather crude
approximation, when it comes to the quantum contributions to the cross
sections, but an inappropriate initial state will spoil even the
classical parts.

The cross sections shown here are by no means exhaustive and at 250 eV
they still have to be confirmed by experiment.  They demonstrate that
it now is possible to perform classical trajectory calculations on
multi electron targets, in this case on helium, and that they yield
reasonable results.  These classical calculations will eventually not
explain every single feature of any arbitrary cross section, but as
they reproduce the classical parts of the reaction's outcome they guide
the identification of the quantum reaction paths and contributions to
the complex low energy dynamics and cross sections.

It should be emphasized that this quasiclassical ansatz is very
similar to standard CTMC and that it is easy to include the initial
backward propagation to stabilize arbitrary unstable initial target 
descriptions into existing CTMC codes --- only the interpretation of 
each single trajectory as one realization of the actual experiment has 
to be discarded. Cross sections are, as in a discretized quantum 
approach, calculated only from the whole set of the trajectories' 
final values.

Further work will focus on the region of very low impact energies to
see how the predicted power law at the threshold will be reproduced
and how good the absolute cross sections coincide with experimental
results. 

\smallskip

This work was funded by the Israel Science Foundation.  The author
wants to thank Alexander Dorn for providing the experimental cross
section of figure \ref{fig:EbEc-Dorn-Cohen}(a) in tabulated form.

\bigskip


\end{document}